\documentclass[12pt]{iopart}

\usepackage{graphicx}

\usepackage{iopams}

\newtheorem{theorem}{Theorem}[section]

\newtheorem{definition}[theorem]{Definition}
\begin{document}
\title[Conics on the Minkowski plane]{Geometry of the conics on the Minkowski plane}
\author{ F. Aceff-S\'anchez $^{1}$\& L. Del Riego Senior$^{2}$}
\address{$^{1}$ Departamento de Matem\'aticas\\
Facultad de Ciencias\\
Universidad Nacional Aut\'onoma de M\'exico}
  \address{ $^{2}$ Departamento de Matem\'aticas\\
Facultad de Ciencias\\
Universidad Aut\'onoma de San Luis Potos\'{\i}}
\eads{\mailto{$^{1}$fmas@hp.fciencias.unam.mx}, \mailto{ $^{2}$ lilia@fciencias.uaslp.mx} }

\begin{abstract}
Conics in the Euclidean space have been known for their geometrical beauty
and also for their power to model several phenomena in real life. It usually happens that when thinking about the conics in a semi-Riemannian manifold,  the equations  and the graphs that come to mind are those of the quadratic Euclidean equations. For example, a circle is always perceived like a closed curve.  We study the geometry of the conics in the semi-Riemannian Minkowski spacetime, and interpret each equation with Euclidean eyes. By defining an extended geometric completeness  for conics, we will show that the conic completeness of   conics can be changed through a Euclidean mirror.
\end{abstract}
\ams{53C50,53A35}
\submitto{\CQG}
\maketitle
\section{Introduction}

All conics are given by an equation in terms of a norm or a distance. In the
Euclidean plane, using the usual distance,  it is possible to classify them as a circle, an ellipse, a
parabola or a hyperbola. 

We wish to extend the notion of geodesic geometrical completeness to conics before proceeding further.

\begin{definition}
A conic curve $\mathcal{C}$ is in\textit{complete} if its maximal domain is
inextendible.
\end{definition}

In the case of the Euclidean manifold  $\mathbb{R}^{2}%
$, completeness of a curve is
equivalent to having its domain convex. The complete conics in this manifold are the ellipses and the parabolas. The hyperbolas are \textit{%
incomplete}.

We will show that the following theorems are valid:
\begin{theorem}
The conic completeness of all conics in the Minkovski spacetime, seen with Euclidean eyes, is not necessarily preserved. The type of conic is not preserved either.
\end{theorem}

In case of the Minkowski ellipse and hyperbola we will also show the following.
 
\begin{theorem}

Let  $\mathbf{x}$ be a point on an ellipse or a hyperbola with focus $\mathbf{p}$ and $\mathbf{q}$ and constant $k; d_{\mathcal{M}}(\mathbf{x},\mathbf{p}) = d_1$ and $d_{\mathcal{M}}(\mathbf{x},\mathbf{q}) = d_2$. The signs of the real numbers $d_1\,\!^2,d_2\,\!^2$ and $k^2$ must be equal.
\end{theorem}

\medskip

Note that the inner product $g_{\mathcal{M}}$ on the tangent manifold of the semi-Riemannian real Minkowski plane $%
({\mathcal{M}},g_{\mathcal{M}})$  is very different from that in a usual Riemannian manifold. For example the Cauchy-Schwartz inequality in not valid anymore, $\sqrt{g(%
\mathbf{x},\mathbf{x})}$ is not a norm neither a seminorm. The
\textquotedblleft length\textquotedblright\ of a non zero vector can be any
real value, not necesarily positive, and there is no natural angular mesure%
\cite{{Beem, Dekster}}.

Distance $d$ is induced by any inner product on a semi-Riemannian manifold in exactly the same manner as in Riemannian manifolds:
\begin{equation}
d(\mathbf{x},\mathbf{y})=\sqrt{g\left( \mathbf{x}-\mathbf{y}%
,\mathbf{x}-\mathbf{y}\right) }.
\end{equation}%
 We are working in real differentiable manifolds, so we will use, when possible, $d^2$.

\section{The Minkowski spacetime}
Let $\mathcal{M}=\mathbb{R}^{2}$ the  manifold with the semi-Riemann metric:
\begin{equation}
g_{\mathcal{M}}(\mathbf{x},\mathbf{y})=x_1 y_1 -x_ 2 y_2.
\end{equation}%

The main calculation in  $\mathcal{M}$ is the following. Let $%
\mathbf{x}=(x,y)$ and $\mathbf{p}=(p_1,p_2)$ be two points
in this space, then 
\begin{equation}
{d_{\mathcal{M}}}^{2}\left( \mathbf{x},\mathbf{p}\right) =\left(
x-p_1\right)^{2}-\left( y - p_2\right)^{2}  \label{distancia}
\end{equation}
\subsection{Ellipses in the Minkowski plane}
An \textit{ellipse} ${\mathcal{E}}$ with focus $\mathbf{p}=(p_1,p_2)$
and $\mathbf{q}=(q_1,q_2)\in \mathcal{M}$ is defined by the set: 
\begin{equation}
{\mathcal{E}}=\{\mathbf{x}=(x,y)\in \mathcal{M}\mid d_{\mathcal{M}%
}(\mathbf{x},\mathbf{p})+ d_{\mathcal{M}}(\mathbf{x},\mathbf{q})=k\},
\label{defelipse}
\end{equation}%
where  $ k^2$ is  a  real number. 

The equation in the reals  of the ellipse ${\mathcal{E}}$ in the Minkowski plane, in terms
of the coordinates of $\mathbf{x}$ and of the focus $\mathbf{p}$ and $\mathbf{q}$, is given by: 

\begin{eqnarray}\label{elipse} 
&  &       
 4 ((q_1- p_1 )^2 -k^2) x^2 +
 8  (q_1 - p_1) \left( p_2 - q_2\right) x y \\ \nonumber 
 && + 4 ( (p_2 - q_2)^2 +  k^2 ) y^2 \\ \nonumber
 && + 4 \left( \left( q_1 - p_1\right) \left(  q_2^{2}- p_1^{2} + q_2^{2}- p_2^{2}- k^{2}
 \right) +  2 k^2 q_1 \right) x\\ \nonumber
 && + 4 \left(\left(  p_2 - q_2\right) \left(q_1^2- p_1^2 + q_2^{2}- p_2^{2} - k^{2}\right) - 2 k^2  q_2 \right)   y\\ \nonumber
  &&  + 
   \left(  q_1^2 - p_1^2 +   q_2^2 - p_2^2 - k^2 \right)^2 - 4 k^{2} q_1^{2} + 4 k^{2} q_2^{2} = 0.\nonumber 
\end{eqnarray}

We have observed that there is a tendency to use the Euclidean conic equations (e.g. Beem (1975) \cite{Beem1} in the case of a hyperbola), when indeed the equations for the conics in $({\mathcal{M}},g_{\mathcal{M}})$ are
different from those in the Euclidean plane. For example, using the Euclidean classification of conics in terms of a general quadratic equation, the discriminant of equation \ref{elipse} is given by
$$ \Delta = 64 k^{2} \left( \left( p_2- q_2\right)^2 - \left( p_1- q_1\right)^2 +  k^2   \right) 
\in \mathbb{R}.$$

 An  ellipse in Minkowski space interpreted as a Euclidean equation corresponds to any kind of conic...  For example the ellipse with $p_2 \neq- q_2 $ with focus  such that  $-\left( p_1 -  q_1\right)^2  + k^2< {(p_2 -q_2)^{2}}$  has 
 $\Delta > 0$, so it is neither complete nor a closed line.
\subsection{Hyperbolas in the Minkowski plane}
A \textit{hyperbola} ${\mathcal{H}}$ with focus $\mathbf{p} =(p_1,p_2)$
and $\mathbf{q}= (q_1,q_2)$ is defined by the set: 
\begin{equation}\label{hyp}
{\mathcal{E}}_{\mathcal M} = \{ \mathbf{x} = (x,y) \in \mathbb{R}^{2} : d_{\mathcal M}\left(%
\mathbf{x},\mathbf{p}\right) - d_{\mathcal{M}}\left(\mathbf{x},\mathbf{q}\right)
= k,\}  
\end{equation}
where $k^{2}$ is a real number.

The real Equation \ref{hyp} of ${\mathcal{H}}$ is  indistinguishable from that of the Minkowski ellipse, so of course  $\Delta $ can have any real value. With a Euclidean differential geometric lens this conic can be seen as a conically incomplete hyperbola   or it is a conically complete ellipse or parabola.
\subsection{Parabolas in the Minkowski plane}
\begin{center}
  \begin{figure}[h]
\begin{picture}(300,220)(0,0)
\put(130,0){\resizebox{8 cm}{!}{\includegraphics{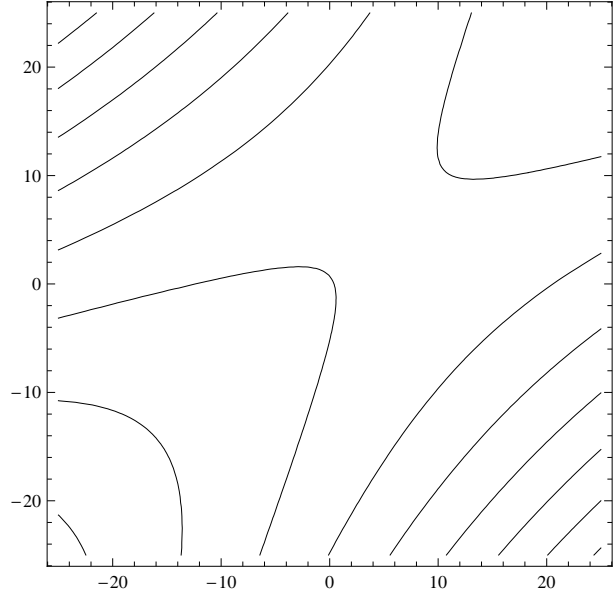}}}
\end{picture}
\caption{The parabola with focus $(2,3)$ and  directrix $(t,2t) \neq 0$,  with  equation $- 3 x^2  - 12 x y - 4 y^2 -36 x - 18 y + 15 = 0$, can be seen in the real plane with a Euclidean lens as conically incomplete. }
\end{figure}
\end{center}

The parabola directrix will be taken as a geodesic $\bgamma:I \subset \mathbb{R} \rightarrow {\mathcal M} $  of ${\mathcal M}$. But geodesics in this manifold are straight lines. 

In order to define a \textit{parabola} ${\mathcal{P}}_{\mathcal M}$ with focus $\mathbf{p} =(p_1,p_2)$ and directrix\\ $\bgamma(t) = ( a t + b, c t + d), a^{2}, b^{2},c^{2},d^{2} \in \mathbb{R} $ we have first to define distance between a point  and a line. We start by finding the critical values   of  the $d_{\mathcal M}\,\!^{2}$ function with variable $t$ 
$$ \left( x - a t - b\right)^2 - \left( y - c t - d\right)^2,$$ 
obtaining 
$$ t_0 = \frac{a x - c y - a b + cd}{a^2-c^2}.$$ We then use this  $t_0$ to obtain a point $ ( a t_0 + b, c t_0 + d)$ in the line $\mathbf{{\mathcal{L}}}$.
\begin{definition}
Let $\mathbf{x}$ be a point  and  $ \mathbf{{\mathcal{L}}}$ be a non nule line, both in ${\mathcal{M}}$. The distance between them is defined by 
\begin{equation*}
d_{\mathcal{M}}\left( \mathbf{x}, \mathbf{{\mathcal{L}}}\right)  = d_{\mathcal{M}}\left(  \mathbf{x}, \left( \frac{a^{2} x - ac y + ac d - c^{2} b}{a^{2}-c^{2}}, \frac{ac x - c^{2} y - a b c + a^{2}d}{a^{2}-c^{2}} \right)\right)  .
\end{equation*}

\end{definition}
 A parabola with focus $\mathbf{p}%
=(p_{1},p_{2})$ and  a non nule directrix  $\mathbf{{\mathcal{L}}}$ can then be defined as the set of points  $\mathbf{x} = (x,y) \in {\mathcal M}$ such that: 
\begin{equation}
 d_{\mathcal M}\left( (x, y), (p_1, p_2)\right)  = d_{\mathcal M}\left( (x, y), \mathbf{{\mathcal{L}}}\right).
\end{equation}

 The real Equation we obtain is the following:
\begin{eqnarray*}
 &&a^{2}\left( a^{2}-c^{2}\right) x^{2}-2ac\left( a^{2}-c^{2}\right)
xy+c^{2}\left( 3a^{2}-c^{2}\right) y^{2} \\
&&+2\left( a^{3}cd-bc^{4}-ac^{3}d-a^{2}bc^{2}-\left( a^{2}-c^{2}\right)^{2}
p_{1}\right) x \\
&&+2\left( \left( a^{2}-c^{2}\right)
p_{2}+a^{3}bc-a^{4}d+a^{2}c^{2}d-abc^{3}\right) y \\
&&+\left( a^{2}-c^{2}\right) \left( p_{1}^{2}-p_{2}^{2}\right)
+a^{2}b^{2}c^{2}+a^{4}d^{2}-2a^{3}bcd-a^{2}c^{2}d^{2}+b^{2}c^{4}+2abc^{3}d = 0
\end{eqnarray*}

  The Euclidean  discriminant $\Delta$ of the parabolas is given by
  $$ 8a^4 c^2(c^2- a^2) \in \mathbb{R}.$$
 
 So the Minkowski parabolas can be seen  with a Euclidean lens as conically complete or conically incomplete. For an example of a conically incomplete parabola, see  Figure~1.

Resuming the situation, someone with Euclidean eyes would see a \textbf{very
different} set of conics in the Minkowski plane:  ellipses may not be even seen as closed curves! Thus Theorem 1.2 is proved.

\begin{center}
 
\begin{tabular}{|c|c|c|}
\hline
 \multicolumn{2}{|c}{Type of Conic}  & \\ \hline\hline
  \textbf{Minkowski Plane} & \textbf{With Euclidean eyes}& \textbf{Conically}\\ \hline
  Circle, Ellipse, Hyperbola, Parabola  & any conic or lines or no conic &   complete  or incomplete\\ \hline
\end{tabular}
\end{center}

Theorem 1.3 is true because to obtain the real equation of the ellipse and hyperbola, which is the same, we must square it twice.


\section*{References}
\begin{thebibliography}{9}
\bibitem{Beem1}  Beem, J 1975
\textit{Geom. Dedicata} \textbf{4}, 45-49.
\bibitem{Beem} Beem, J, Ehrlich, \thinspace\ P and Easley, K 1981
Global Lorentzian Geometry, \textit{Dekker}.
\bibitem{Dekster} Dekster, B, \textit{J.
Geom.} \textbf{80} (2004) 31-47.

\bibitem{Dodson} Dodson C and Poston T 1980 Tensor Geometry, The
Geometric Viewpoint and its uses, Second Edition, \textit{Springer-Verlag}.

\bibitem{O'Neil1} O'Neill B 1997\\ Elementary Differential Geometry.\\ \textit{Academic Press}.

\bibitem{O'Neil} O'Neill B  1983 \\ Semi-Riemannian geometry, with
applications to Relativity\\ \textit{Elsevier Science}.

\bibitem{Lang} Lang S 1962 \\ Differential and Riemannian manifolds\\
Graduate Texts in Mathematics \textbf{160}, \textit{Springer Verlag}.

\bibitem{Stewart} Stewart, J 1990 Advanced general relativity, %
\textit{Cambridge University Press}.
\end{thebibliography}
\end{document}